\documentclass[preprint,showpacs,preprintnumbers,amsmath,amssymb,natbib]{revtex4}

\usepackage{graphicx}
\usepackage{epsfig}		
\usepackage{dcolumn}
\usepackage{bm}
\def\0{\mbox{\tiny $0$}}
\def\1{\mbox{\tiny $1$}}
\def\2{\mbox{\tiny $2$}}
\def\3{\mbox{\tiny $3$}}
\def\4{\mbox{\tiny $4$}}
\def\5{\mbox{\tiny $5$}}
\def\6{\mbox{\tiny $6$}}
\def\7{\mbox{\tiny $7$}}
\def\8{\mbox{\tiny $8$}}
\def\9{\mbox{\tiny $9$}}

\def\f14{\mbox{\tiny $\frac{1}{4}$}}

\def\ii{\mbox{\tiny $i$}}

\def\s{\mbox{\tiny $s$}}

\def\j{\mbox{\tiny $j$}}
\def\mi{\mbox{\tiny $-$}}

\def\pl{\mbox{\tiny $+$}}
\def\al{\mbox{\tiny $\alpha$}}

\def\bb#1{\mbox{\footnotesize $(#1)$}}

\begin{document}

\title{Dynamical neutrino masses in the generalized Chaplygin gas scenario with mass varying CDM}

\author{A. E. Bernardini}
\affiliation{Departamento de F\'{\i}sica, Universidade Federal de S\~ao Carlos, PO Box 676, 13565-905, S\~ao Carlos, SP, Brasil}
\email{alexeb@ufscar.br, alexeb@ifi.unicamp.br}

\date{\today}

\begin{abstract}
Neutrinos coupled to an underlying scalar field in the scenario for unification of mass varying dark matter and cosmon-{\em like} dark energy is examined.
In the presence of a tiny component of mass varying neutrinos, the conditions for the present cosmic acceleration and for the stability issue are reproduced.
It is assumed that {\em sterile} neutrinos behave like mass varying dark matter coupled to mass varying {\em active} neutrinos through the {\em seesaw} mechanism, in a kind of {\em mixed} dark matter sector.
The crucial point is that the dark matter mass may also exhibit a dynamical behavior driven by the scalar field.
The scalar field mediates the nontrivial coupling between the mixed dark matter and the dark energy responsible for the accelerated expansion of the universe.
The equation of state of perturbations reproduce the generalized Chaplygin gas (GCG) cosmology so that all the effective results from the GCG paradigm are maintained, being perturbatively modified by neutrinos.
\end{abstract}

\date{\today}
\maketitle

\section{Introduction}

The continuous search for the physical mechanism which sets the cosmic acceleration of the universe and the corresponding conditions for stability has stimulated interesting and sometimes fascinating discussions on cosmological models \cite{Zla98,Wan99,Ste99,Bar99,Ber00}.
The dynamical mass attributed to neutrinos or dark matter \cite{Bea08,novel} can, for instance, regulate the time evolution of the dynamical dark energy providing the setup of the cosmic acceleration followed by the cosmological stability.
In this context, the coupling of mass varying dark matter with neutrinos yields interesting relations between the present mass of neutrinos and the dark energy equation of state.

In a previous issue \cite{Ber09}, it was demonstrated that an effective generalized Chaplygin gas (GCG) scenario \cite{Kam02,Bil02,Ber02} can be reproduced in terms of a dynamical dark energy component $\phi$ with equation of state given by $p_{\phi}\bb{\phi} = -\rho_{\phi}\bb{\phi}$ and a cold dark matter (CDM) component with a dynamical mass driven by the scalar field $\phi$.
Dark matter is, most often, not considered in the mass varying neutrino (MaVaN) models.
The treatment of dark energy and dark matter in the GCG unified scheme naturally offers this possibility.
Identifying sterile neutrinos as dark matter coupled with dark energy provides the conditions to implement such unified picture in the MaVaN formulation since active and sterile neutrino states are connected through the {\em seesaw} mechanism for mass generation.
The constraints imposed by the {\em seesaw} mechanism allows one to establish an analytical connection to the GCG in terms of a real scalar field.
The dynamics of the coupled fluid composed by neutrinos, dark matter and dark energy is driven by one single degree of freedom, the scalar field, $\phi\bb{a}$.

The simplest realization of MaVaN mechanisms \cite{Hun00,Gu03,Far04,Pec05,Ber08A} consists in writing down an effective potential which, in addition to a scalar field dependent term, contains a term related to the neutrino energy density.
It results in the so-called adiabatic condition, which sometimes leads to the stationary regime for the scalar field in respect with an effective potential \cite{Far04,Bro06A,Bja08,Ber08B}.
One indeed expect a tiny contribution from cosmological neutrinos to the energy density of the universe.
MaVaN scenarios essentially predict massless neutrinos until recent times.
When their mass eventually grows close to its present value, they form a non-relativistic (NR) fluid and the interaction with the scalar field stops its evolution.
The relic particle mass is generated from the vacuum expectation value of the scalar field and becomes linked to its dynamics: $m\bb{\phi}$.
It is presumed that the neutrino mass has its origin on the vacuum expectation value (VEV) of the scalar field and its behavior is governed by the dependence of the scalar field on the scale factor.

In fact, it is well-known that the active neutrino masses are tiny as compared to the masses of the charged fermions.
This can be understood through the symmetry of the standard model (SM) of electroweak interactions.
It involves only left-handed neutrinos such that no renormalizable mass term for the neutrinos is compatible with the SM gauge symmetry $SU(2)_{L}\otimes U(1)_{Y}$.
Once one has assumed that baryon number and lepton number is conserved for renormalizable interactions, neutrino masses can only arise from an effective dimension five operator.
It involves two powers of the vacuum expectation value of the Higgs doublet.
They are suppressed by the inverse power of a large mass scale $M$ of a sterile right-handed Majorana neutrino, since it has hypercharge null in the SM.
This super-massive Majorana neutrino should be characteristic for lepton number violating effects within possible extensions beyond the SM.
At our approach, the mass scale $M$ has its dynamical behavior driven by $\phi$.
In some other words, the sterile neutrino mass is characteristic of the aforementioned mass varying dark matter which indirectly results in MaVaN's, i. e. active neutrinos with mass computed through {\em seesaw} mechanism.

In section II, we report about the main properties of a unified treatment of dark matter and dark energy prescribed by the mass varying mechanism.
One sees that model dependent choices of dynamical masses of dark matter allows for reproducing the conditions for the present cosmic acceleration in an effective GCG scenario.
The stability conditions resulted from a positive squared speed of sound, $c_{s}^{\2}$, is recovered.
In section III, we discuss the neutrino mass generation mechanism in the context of the GCG model.
Following the simplest formulation of the {\em seesaw} mechanism, Dirac neutrino masses with analytical dependencies on $\phi$, $1$ and $\phi^{\mi\1}$ are considered.
In section IV, we discuss the conditions for stability and the perturbative modifications on the accelerated expansion of the universe in the framework here proposed.
One can state the conditions for reproducing the GCG scenario.
We draw our conclusions in section V.

\section{Mass varying dark matter and the GCG}

To understand how the mass varying mechanism takes place for different particle species, it is convenient to describe the relevant physical variables as functionals of a statistical distribution $f\bb{q}$.
This counts the number of particles in a given region around a point of the phase space defined by the conjugate coordinates: momentum, $\beta$, and position, $x$.
The statistical distribution $f\bb{q}$ can be defined in terms of the comoving momentum, $q$, and for the case where $f\bb{q}$ is a Fermi-Dirac distribution, it can be written as
\begin{equation}
f\bb{q}= \left\{\exp{\left[(q - q_F)/T_{\0}\right]} + 1\right\}^{\mi\1},
\end{equation}
where $T_{\0}$ is the relic particle background temperature at present.

In the flat FRW scenario, the corresponding particle density, energy density and pressure can thus be depicted from the Einstein's energy-momentum tensor \cite{Dod05} as
\begin{eqnarray}
n\bb{a} &=&\frac{1}{\pi^{\2}\,a^{\3}}
\int_{_{0}}^{^{\infty}}{\hspace{-0.3cm}dq\,q^{\2}\ \hspace{-0.1cm}f\bb{q}},\nonumber\\
\rho_M\bb{a, \phi} &=&\frac{1}{\pi^{\2}\,a^{\4}}
\int_{_{0}}^{^{\infty}}{\hspace{-0.3cm}dq\,q^{\2}\, \left(q^{\2}+ M^{\2}\bb{\phi}\,a^{\2}\right)^{\1/\2}\hspace{-0.1cm}f\bb{q}},\\
p_M\bb{a, \phi} &=&\frac{1}{3\pi^{\2}\,a^{\4}}\int_{_{0}}^{^{\infty}}{\hspace{-0.3cm}dq\,q^{\4}\, \left(q^{\2}+ M^{\2}\bb{\phi}\,a^{\2}\right)^{\mi\1/\2}\hspace{-0.1cm} f\bb{q}}.~~~~ \nonumber
\label{gcg01}
\end{eqnarray}
where $a$ is the scale factor (cosmological radius) for the flat FRW universe, for which the metrics is given by $ds^{\2} = dt^{\2} - a^{\2}\bb{t}\delta_{\ii\j}dx^{\ii}dx^{\j}$.

From the dependence of $\rho$ on $a$ one can obtain the energy-momentum conservation equation,
\begin{equation}
\dot{\rho_M} + 3 H (\rho_M + p_M) - \dot{\phi}\frac{\mbox{d} M}{\mbox{d} \phi} \frac{\partial \rho_M}{\partial M} = 0,
\label{gcg03}
\end{equation}
that translates the dependence of $M$ on $\phi$ into a dynamical behavior, and where $H = \dot{a}/{a}$ is the expansion rate of the universe and the {\em overdot} denotes differentiation with respect to time ($^{\cdot}\, \equiv\, d/dt$).
Simple mathematical manipulations allow one to easily demonstrate that
\begin{equation}
M\bb{a} \frac{\partial \rho_M\bb{a}}{\partial M\bb{a}} = (\rho_M\bb{a} - 3 p_M\bb{a}),
\label{gcg02}
\end{equation}
from which, one can see that the strength of the coupling between relic particles and the scalar field is suppressed by the relativistic increase of pressure ($\rho\sim 3 p$).
As long as particles become ultra-relativistic ($T\bb{a} = T_{\0}/a \propto q/a >> M\bb{\phi\bb{a}}$) the matter fluid and the scalar field fluid tend to decouple and evolve adiabatically separated.

Adding Eq.~(\ref{gcg03}) to the equation of motion for a cosmon field $\phi$ like
\begin{equation}
\ddot{\phi} + 3 H \dot{\phi} + \frac{\mbox{d} U\bb{\phi}}{\mbox{d} \phi} =  Q\bb{\phi} = - \frac{\mbox{d} M}{\mbox{d} \phi} \frac{\partial \rho_M}{\partial M},
\label{gcg04}
\end{equation}
rewritten as
\begin{equation}
\dot{\rho_{\phi}} + 3 H (\rho_{\phi} + p_{\phi}) + \dot{\phi}\frac{\mbox{d} M}{\mbox{d} \phi} \frac{\partial \rho_M}{\partial M} = 0,
\label{gcg05}
\end{equation}
results in the equation of energy conservation for a unified fluid with a dark energy component and
a mass varying dark matter component,
\begin{equation}
\dot{\rho} + 3 H (\rho + p) = 0.
\label{gcg06}
\end{equation}
for which we identify $\rho = \rho_{\phi} + \rho_M$ and $p = p_{\phi} + p_M$.

From this point, it is suggested that such a unified fluid corresponds to an effective description of a GCG universe.
The GCG model is characterized by an exotic equation of state \cite{Ber02,Ber03} given by
\begin{equation}
p = - A_{\s} \rho_{\0} \left(\frac{\rho_{\0}}{\rho}\right)^{\al},
\label{gcg20}
\end{equation}
which can be obtained from a generalized Born-Infeld action \cite{Ber02}.
Inserting the above equation of state into the unperturbed energy conservation Eq.~(\ref{gcg06}) and following a straightforward integration \cite{Ber02}, one obtains
\begin{equation}
\rho = \rho_{\0} \left[A_{\s} + \frac{(1-A_{\s})}{a^{\3(\1+\alpha)}}\right]^{\1/(\1 \pl \al)},
\label{gcg21}
\end{equation}
and
\begin{equation}
p = - A_{\s} \rho_{\0} \left[A_{\s} + \frac{(1-A_{\s})}{a^{\3(\1+\alpha)}}\right]^{-\al/(\1 \pl \al)}.
\label{gcg22}
\end{equation}

One of the most striking features of the GCG fluid is that its energy density interpolates between a dust dominated phase, $\rho \propto a^{-\3}$, in the past, and a de-Sitter phase, $\rho = -p$, at late times.
This property makes the GCG model an interesting candidate for the unification of dark matter and dark energy.
Notice that for $A_s =0$, GCG behaves always as matter whereas for $A_{\s} =1$, it behaves always as a cosmological constant.
Hence to use it as a unified candidate for dark matter and dark energy one has to exclude these two possibilities so that $A_s$ must lie in the range $0 < A_{\s} < 1$.
Furthermore, this evolution is controlled by the model parameter $\alpha$.

Assuming the canonical parametrization of $\rho$ and $p$ in terms of a scalar field $\phi$,
\begin{eqnarray}
\rho &=& \frac{1}{2}\dot{\phi}^{\2} + V\bb{\phi},\nonumber\\
p    &=& \frac{1}{2}\dot{\phi}^{\2} - V\bb{\phi},
\label{pap01}
\end{eqnarray}
and following Ref.~\cite{Ber04}, one can obtain through Eq.~(\ref{gcg21}-\ref{gcg22}) the effective dependence of $\phi$ on $a$ implicitly given by
\begin{equation}
\dot{\phi}^{\2}\bb{a} = \frac{\rho_{\0}(1 - A_{\s})}{a^{\3(\al\pl\1)}}
\left[A_{\s} + \frac{(1-A_{\s})}{a^{\3(\al\pl\1)}}\right]^{-\al/(\al \pl \1)},
\label{pap02}
\end{equation}
and explicit expressions for $\rho$, $p$ and $V$ in terms of $\phi$.
Assuming a flat evolving universe described by the Friedmann equation $H^{\2} = \rho$ (with $H$ in units of $H_{\0}$ and $\rho$ in units of $\rho_{\mbox{\tiny Crit}} = 3 H^{\2}_{\0}/ 8 \pi G)$, one obtains
\begin{equation}
\phi\bb{a} = - \frac{1}{3(\alpha + 1)}\ln{\left[\frac{\sqrt{1 - A_{\s}(1 - a^{\3(\al \pl \1)})} - \sqrt{1 - A_{\s}}}{\sqrt{1 - A_{\s}(1 - a^{\3(\al \pl \1)})} + \sqrt{1 - A_{\s}}}\right]},
\label{pap03}
\end{equation}
where it is assumed that
\begin{equation}
\phi_{\0} = \phi\bb{a_{\0} = 1} = - \frac{1}{3(\alpha + 1)}\ln{\left[\frac{1 - \sqrt{1 - A_{\s}}}{1 + \sqrt{1 - A_{\s}}}\right]}.
\label{pap04}
\end{equation}
One then readily finds the scalar field potential,
\begin{equation}
V\bb{\phi} = \frac{1}{2}A_{\s}^{\frac{\1}{\1 \pl \al}}\rho_{\0}\left\{
\left[\cosh{\left(3\bb{\alpha + 1} \phi/2\right)}\right]^{\frac{\2}{\al \pl \1}}
+
\left[\cosh{\left(3\bb{\alpha + 1} \phi/2\right)}\right]^{-\frac{\2\al}{\al \pl \1}}
\right\}.
\label{pap05}
\end{equation}

If one supposes that the energy density, $\rho$, may be decomposed into a mass varying CDM component, $\rho_{M}$, and a dark energy component, $\rho_{\phi}$, connected by the scalar field equations (\ref{gcg04})-(\ref{gcg05}), the equation of state (\ref{gcg20}) is just assumed as an effective description of the cosmological background fluid of the universe.
Since the CDM pressure, $p_M$, is null, the dark energy component of pressure, $p_{\phi}$, results in the GCG pressure, $p = p_{\phi}$.
Assuming that dark energy behaves like a cosmological constant, that is, its equation of state is given by $\rho_{\phi}\bb{\phi} = - p_{\phi}\bb{\phi}$, the dark energy density can be parameterized by a generic quintessence potential, $\rho_{\phi}\bb{\phi} = U\bb{\phi}$, since its kinetic component has to be null for a canonical formulation.
It results in $U\bb{\phi} = - p_{\phi}\bb{\phi} = p$, where $p$ is the GCG pressure given by Eq.~(\ref{gcg22}).
By substituting the result of Eq.(\ref{pap03}) into the Eq.(\ref{gcg22}), and observing that $H^{\2} = \rho$, with $\rho$ given by Eq.(\ref{gcg21}), it is possible to rewrite the GCG pressure, $p$, in terms of $\phi$.
It results in the following analytical expression for $U\bb{\phi}$,
\begin{equation}
U\bb{\phi} = \rho_{\phi}\bb{\phi} = - p_{\phi}\bb{\phi} = \left[A_{s}\cosh{\left(\frac{3\bb{\alpha + 1}\phi}{2}\right)}\right]^{-\frac{\2 \al}{1 \pl \al}},
\label{pap08}
\end{equation}
which is consistent with the result for $V\bb{\phi} = (1/2)(\rho\bb{\phi} - p\bb{\phi})$ from Eq.~(\ref{pap05}).
Since $\rho_{\phi}\bb{\phi} + p_{\phi}\bb{\phi} = 0$, the Eq.~(\ref{gcg05}) is thus reduced to
\begin{equation}
\frac{\mbox{d} U\bb{\phi}}{\mbox{d}{\phi}} +
\frac{\partial \rho_{M}}{\partial M} \frac{\mbox{d} M\bb{\phi}}{\mbox{d}\phi} = 0,
\label{pap09}
\end{equation}
and the problem is then reduced to finding a relation between the scalar potential $U\bb{\phi}$ and the variable mass $M\bb{\phi}$.

From the above equation, the effective potential governing the evolution of the scalar field is naturally decomposed into a sum of two terms, one arising from the original quintessence potential $U\bb{\phi}$, and other from the dynamical mass $M\bb{\phi}$.
For appropriate choices of potentials and coupling functions satisfying Eq.~(\ref{pap09}), the competition between these terms leads to a minimum of the effective potential.
In the adiabatic regime, the matter and the scalar field are tightly coupled together and evolve as one effective fluid.
At our approach, once one assumes a $\Lambda$-{\em like} equation of state $p_{\phi} = -\rho_{\phi}$, without any additional constraint on cosmon field equations, Eq.~(\ref{pap09}) is naturally obtained.
In the GCG cosmological scenario, the effective fluid description is valid for the background cosmology and for linear perturbations.
The equation of state of perturbations is the same as that of the background cosmology where all the effective results of the GCG paradigm are maintained.

The Eq.~(\ref{pap08}) leads to $\rho + p = \rho_{M} + p_{M}$ which, in the CDM limit, gives
\begin{equation}
\rho\bb{a} + p\bb{a} = M\bb{a} \, n\bb{a} (+ p_{M} \equiv 0).
\label{pap10}
\end{equation}
Since the dependence of $M$ on $a$ is exclusively intermediated by $\phi\bb{a}$, i. e. $M\bb{a} \equiv M\bb{\phi\bb{a}}$, from Eqs.~(\ref{gcg21}), (\ref{gcg22}) and (\ref{pap03}), after simple mathematical manipulations, one obtains
\begin{equation}
M\bb{\phi} = M_{\0} \left[\frac{\tanh{\left(3\bb{\alpha + 1}\frac{\phi}{2}\right)}}{\tanh{\left(3\bb{\alpha + 1}\frac{\phi_{\0}}{2}\right)}}\right]^{\frac{ \2 \al}{1 \pl \al}}
\label{pap11}
\end{equation}
which is consistent with Eq.~(\ref{pap09}).
From the above result, one can infers that the adequacy to the adiabatic regime is left to the mass varying mechanism which drives the cosmological evolution of the dark matter component.
For mass varying CDM coupled with $\Lambda$-{\em like} dark energy, with $p_{\phi} = -\rho_{\phi}$, the GCC leads to similar predictions for the equation of state, $w = p/\rho$, independently of the scale parameter $a$.
The same is not true for hot dark matter (HDM) \cite{Ber09}.

\section{Perturbative approach for neutrino masses}

Unlike photons and baryons, cosmological neutrinos have not been observed, so arguments about their contribution to the total energy density of the universe are necessarily theoretical.
Otherwise, neutrinos coupled to dark energy can lead to a number of significant phenomenological consequences.
The neutrino mass should be a dynamical quantity and neutrinos remain essentially massless until recent times.
When their mass eventually increases close to its present value, their interaction with the background scalar field almost ceases \cite{Ber08A,Ber08B}.
The energy of the scalar field becomes the dominant contribution to the energy density of the Universe and cosmic acceleration ensues.

Without loss of generality, such a behavior can be easily implemented to a degenerate fermion gas (DFG) of neutrinos.
In the limit where $T_{\0}$ tends to $0$ in Eq.~(\ref{gcg01}), the Fermi distribution $f\bb{q}$ becomes a step function that yields an elementary integral for the above equations, with the upper limit equal to the Fermi momentum here written as $\beta\bb{a} = q_{F}/a$.
It results in the equations for a DFG \cite{ZelXX}, which can be useful for parameterizing the transition between ultra-relativistic (UR) and non-relativistic (NR) thermodynamic regimes.
The equation of state can be expressed in terms of elementary functions of $\beta \equiv \beta\bb{a}$ and $m \equiv m\bb{\phi\bb{a}}$,
\begin{eqnarray}
n\bb{a} &=& \frac{1}{3 \pi^{\2}} \beta^{\3}\nonumber\\
\rho_m\bb{a} &=& \frac{1}{8 \pi^{\2}}
\left[\beta(2 \beta^{\2} + m^{\2})\sqrt{\beta^{\2} + m^{\2}} -
\mbox{arc}\sinh{\left(\beta/m\right)}\right]\\
p_m\bb{a} &=& \frac{1}{8 \pi^{\2}}
\left[\beta (\frac{2}{3} \beta^{\2} - m^{\2})\sqrt{\beta^{\2} + m^{\2}} + \mbox{arc}\sinh{\left(\beta/m\right)}\right]\nonumber
\label{gcg01D}
\end{eqnarray}
The relation given by Eq.~(\ref{gcg02}) can be verified for the above definition.
Besides the mass dependence, it is necessary to determine for which values of the scale factor the neutrino-scalar field coupling becomes important.
The DFG approach take care of this.

Let us then turn back to the explanation for the smallness of the neutrino masses.
According to the {\em seesaw} mechanism, the tiny masses, $m$, of the usual left-handed neutrinos are obtained through a very massive, $M$, sterile right-handed neutrino.
The Lagrangian density that describes the simplest version of the {\em seesaw} mechanism via Yukawa coupling between a light scalar field and a single neutrino flavor is given by
\begin{equation}
{\cal L} = m_{LR} \bar{\nu}_L \nu_R + M\bb{\phi} \bar{\nu}_R \nu_R + h.c.,
\label{gcg30}
\end{equation}
where it is shown that at scales well below the right-handed neutrino mass, one has the effective Lagrangian density \cite{seesaw}
\begin{equation}
{\cal L} = \frac{m_{LR}^{\2}}{M\bb{\phi}} \bar{\nu}_L \nu_R + h.c..
\label{gcg31}
\end{equation}
Phenomenological consistency with the SM implies that logarithm corrections to the above terms are small, while
it is well-know from the results of solar, atmospheric, reactor and accelerator neutrino oscillation experiments that neutrino masses given by $m \sim m_{LR}^{\2}\bb{\phi}/M\bb{\phi}$ lie in the sub-$eV$ range.
It is also clear that promoting the scalar field $\phi$ into a dynamical quantity leads to a mechanism in the context of which neutrino masses are time-dependent.
Associating the scalar field to the dark energy field allows linking NR neutrino energy densities to late cosmological times \cite{Far04,Bro06A,Bja08}.
This scenario can be implemented through the perturbative approach via Eq.~(\ref{gcg06}) \cite{Ber09}.
It is evident that this approach is fairly general, as well as independent of the choice of the equation of state and of the dependence of the neutrino mass on the scalar field.

\begin{figure}
\vspace{-3 cm}
\centerline{\psfig{file= 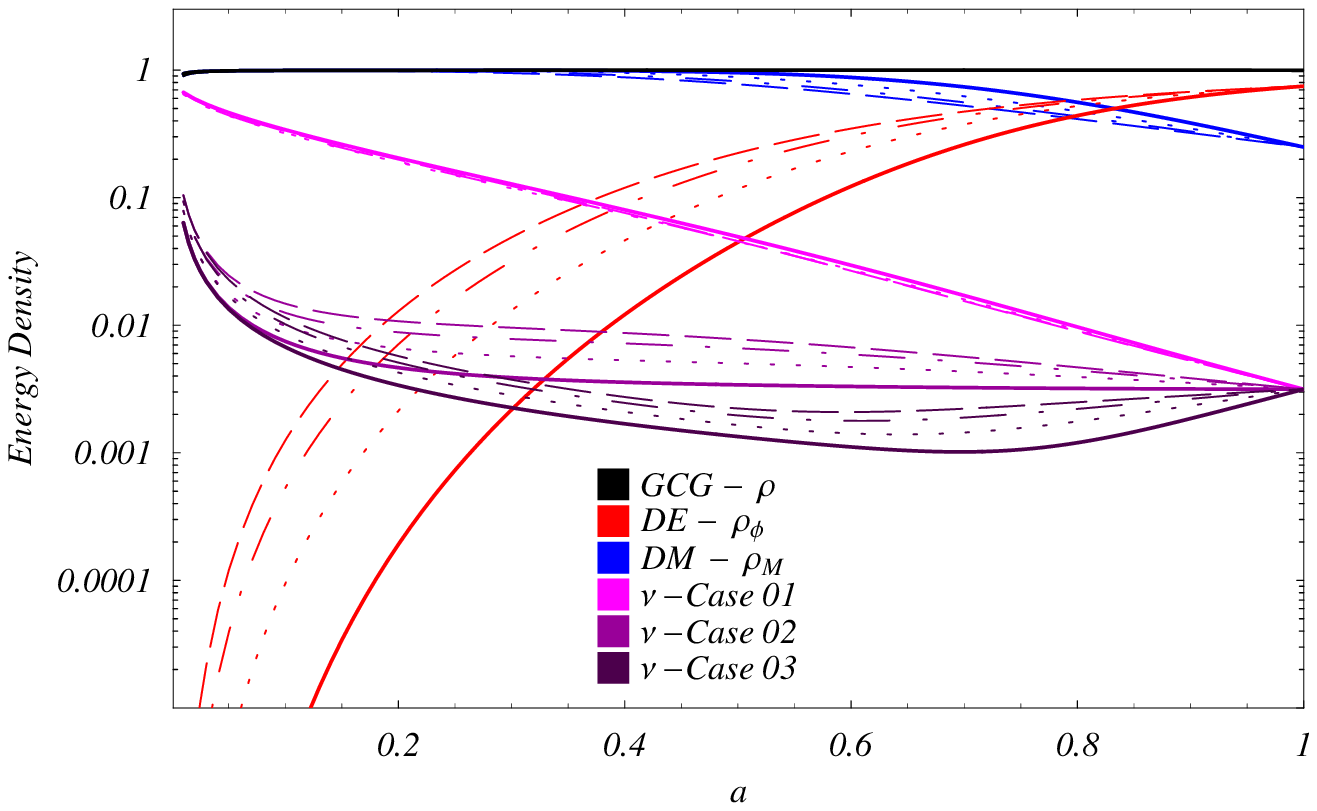,width=14 cm}}
\vspace{-2 cm}
\caption{\small Energy densities as function of the scale factor for the GCG fluid (GCG) and for the decomposed components of the unified background fluid (effective GCG): mass varying dark matter (DM), $\rho_{M}$, cosmon-{\em like} dark energy (DE), $\rho_{\phi}$, and perturbative DFG of neutrinos, $\rho_m$.
For the 03 cases of MaVaN's, the dynamical mass dependencies on $\phi$ are: $\phi^{\2}/M\bb{\phi}$,  $1/M\bb{\phi}$ and $\phi^{\mi\2}/M\bb{\phi}$.
Neutrino densities are computed for a mass varying behavior of $M\bb{\phi}$ in correspondence to a GCG scenario with $A_{\s}= 3/4$ and $\alpha = 1$ (solid line), $1/2$ (dot line), $1/4$ (dash dot line), and $1/8$ (dash line).}
\label{GCG-0}
\end{figure}

\begin{figure}
\vspace{-3 cm}
\centerline{\psfig{file= 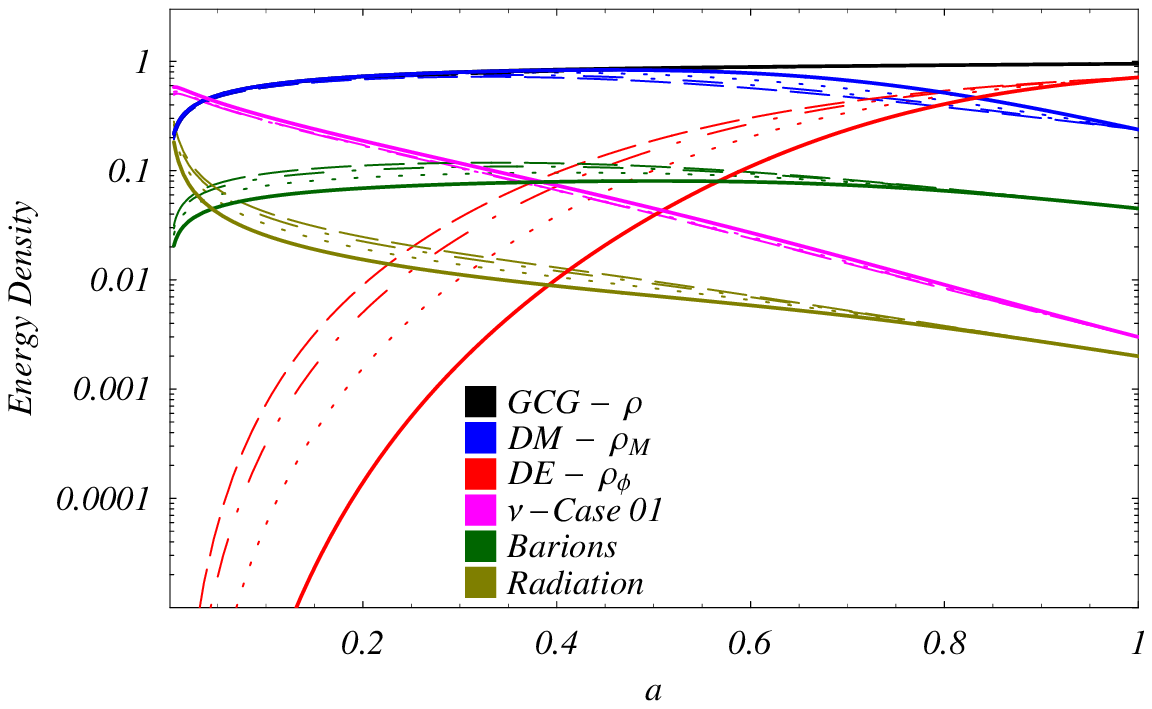,width=10 cm}}
\vspace{-3 cm}
\centerline{\psfig{file= 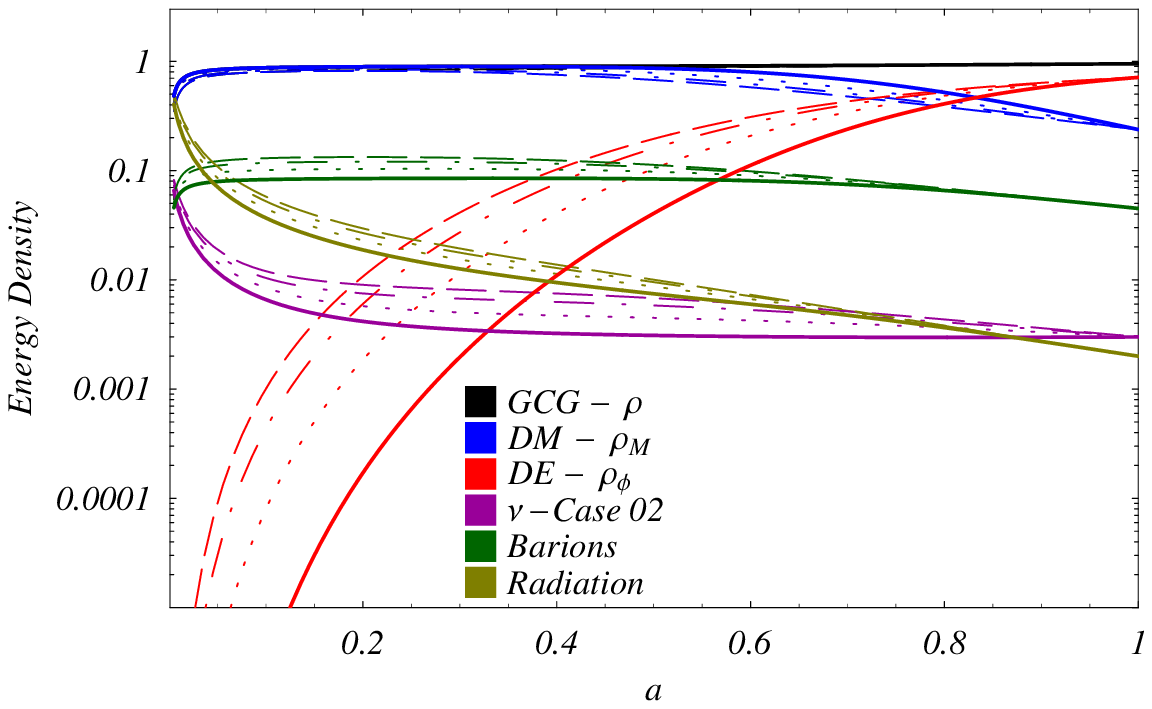,width=10 cm}}
\vspace{-3 cm}
\centerline{\psfig{file= 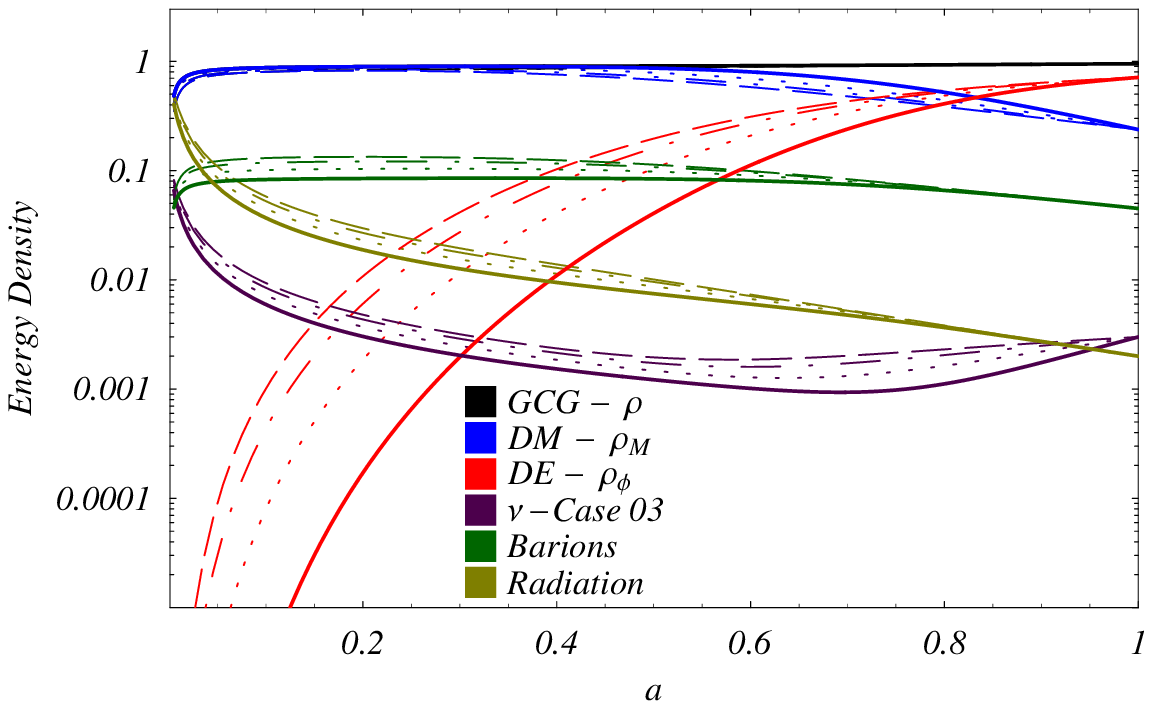,width=10 cm}}
\vspace{-2 cm}
\caption{\small Energy densities ($\rho/\rho_{\mbox{\tiny Crit}}$) as function of the scale factor for all energy components of the universe.
The results are compared with those of an exact GCG scenario with $A_{\s}= 3/4$ and parameters $\alpha = 1$ (solid line), $1/2$ (dot line), $1/4$ (dash dot line), and $1/8$ (dash line).
We have considered all the 03 cases for MaVaN's in correspondence with Fig.~\ref{GCG-0}.}
\label{GCG-B}
\end{figure}

The form of $M\bb{\phi}$ and of the equation of state can indeed lead to quite different scenarios.
In particular, we consider three cases of neutrino mass dependence on $\phi$.
In case 01 we set $m \propto \phi^{\2}/M\bb{\phi}$, in case 02 we have set $m \propto 1/M\bb{\phi}$, and in case 03 we have set $m \propto \phi^{\mi\2}/M\bb{\phi}$.
Energy densities, $\rho$, as function of the scale factor for the GCG fluid and for the decomposed components of the unified effective GCG background fluid, namely, the mass varying dark matter, $\rho_{m}$, the cosmon-{\em like} dark energy, $\rho_{\phi}$, and the perturbative DFG of neutrinos, $\rho_\nu$, are obtained in the Fig.~\ref{GCG-0}.
The neutrino densities are computed for neutrinos masses correlated to the dark matter mass, $M\bb{\phi}$, in correspondence with the GCG scenario with $A_{\s}= 3/4$, that represents a realistic fraction of dark energy, and with $\alpha = 1, \, 1/2, \, 1/4$ and $1/8$.
In the Fig.~\ref{GCG-B} we have taken into account the energy densities for photons ($\rho_{\gamma}\bb{a} = \rho_{\gamma\0} \, a^{\mi\4}$) and baryons ($\rho_{B}\bb{a} = \rho_{B\0} \, a^{\mi\3}$) in complement to our effective GCG scenario.
Then we have computed the relative modifications on the spectrum of the cosmic evolution of the dark sector components.

Conveniently in agreement with the phenomenological predictions, we have set the present values of the energy densities as: $\rho_{\phi\0} = 0.95 A_{\s}$, $\rho_{m\0} = 0.95 (1 - A_{\s})$, $\rho_{\nu\0} = 0.003$, $\rho_{\gamma\0} = 0.002$, and $\rho_{B\0} = 0.045$.
Assuming that neutrinos are NR at present, and observing the behavior of Eqs.~(\ref{gcg01D}), we have set $\beta\bb{a_{\0}} = \beta\bb{1} = \rho_{\nu\0}^{\frac{\1}{\3}}$.
This condition allows one to establish the correspondence between the values of $a$ for which the coupling transition takes place, i. e. the transition between relativistic and NR regimes, so that the neutrino masses achieve the present-day values.
For our model such a transition is governed by the DFG behavior of Eq.~(\ref{gcg01D}).

These scenarios are by no means the only possibilities.
In particular, we choose them as they correspond to the simplest feasible possibility for active neutrino mass generation, given that the scalar field has mass dimension one \cite{seesaw}.
The essential information of the mass dependence on the scale factor is that, for case 03, the neutrino mass increases with decreasing $\phi$, in opposition to what happens for cases 01 and 02.
Through the stability analysis performed in the next section we shall elucidate some important consequences due to such a distinct behavior.

\begin{figure}
\vspace{-3 cm}
\centerline{\psfig{file= 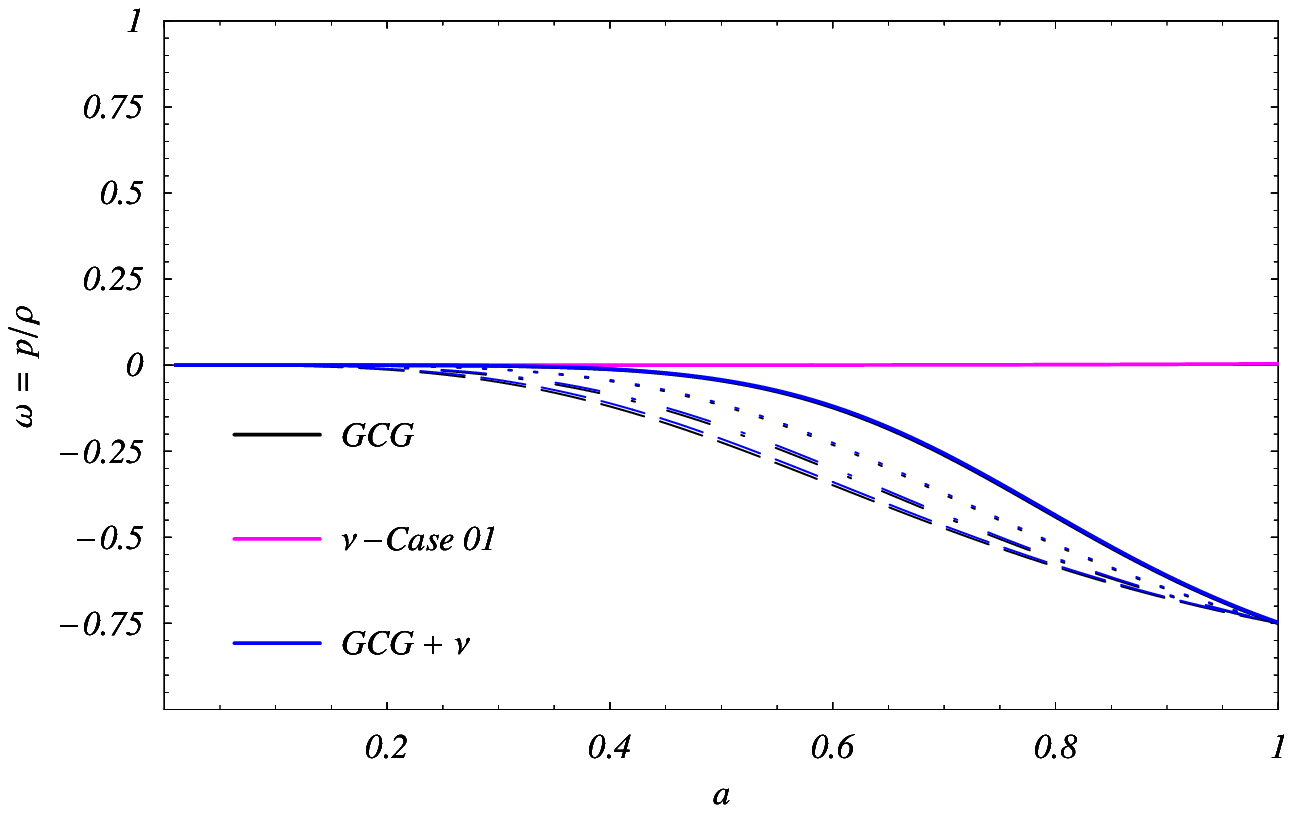,width=9.8 cm}}
\vspace{-4 cm}
\centerline{\psfig{file= 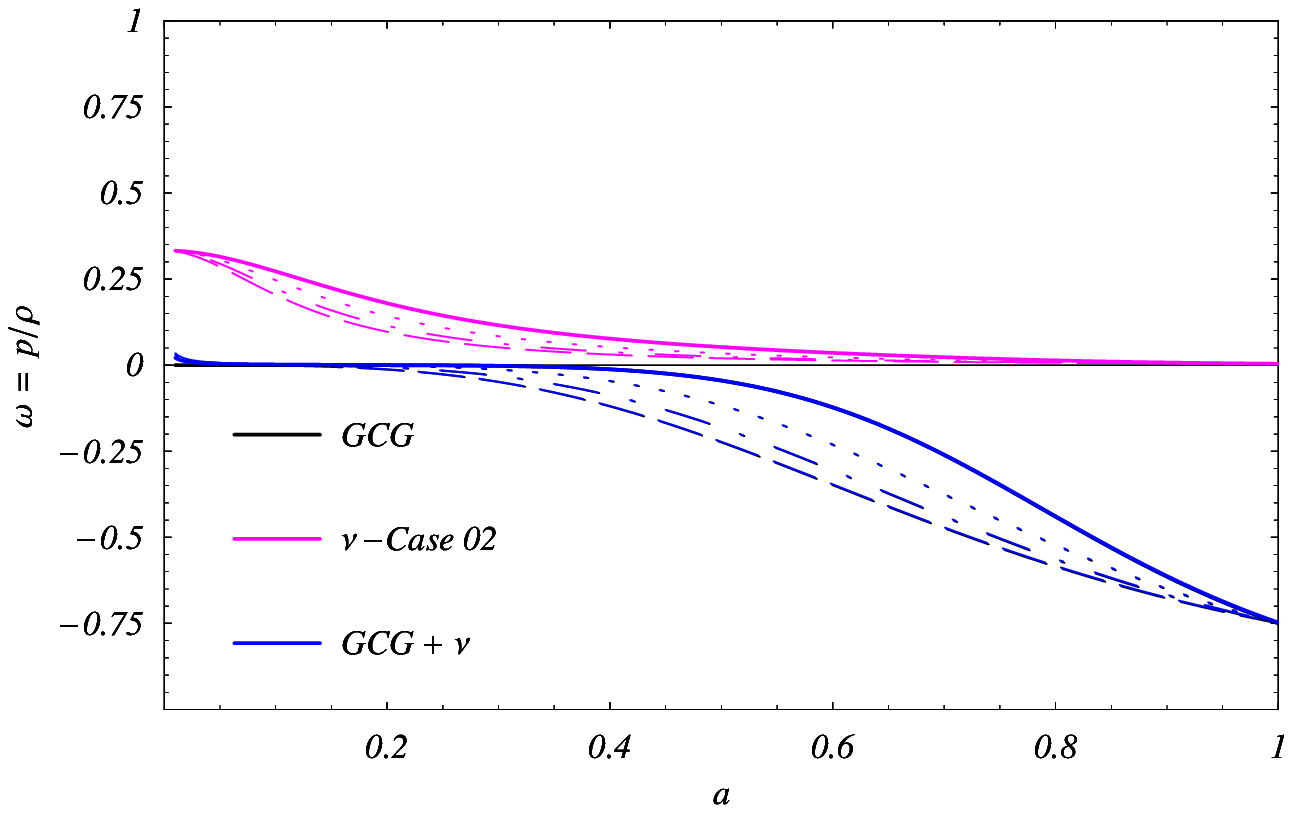,width=9.8 cm}}
\vspace{-4 cm}
\centerline{\psfig{file= 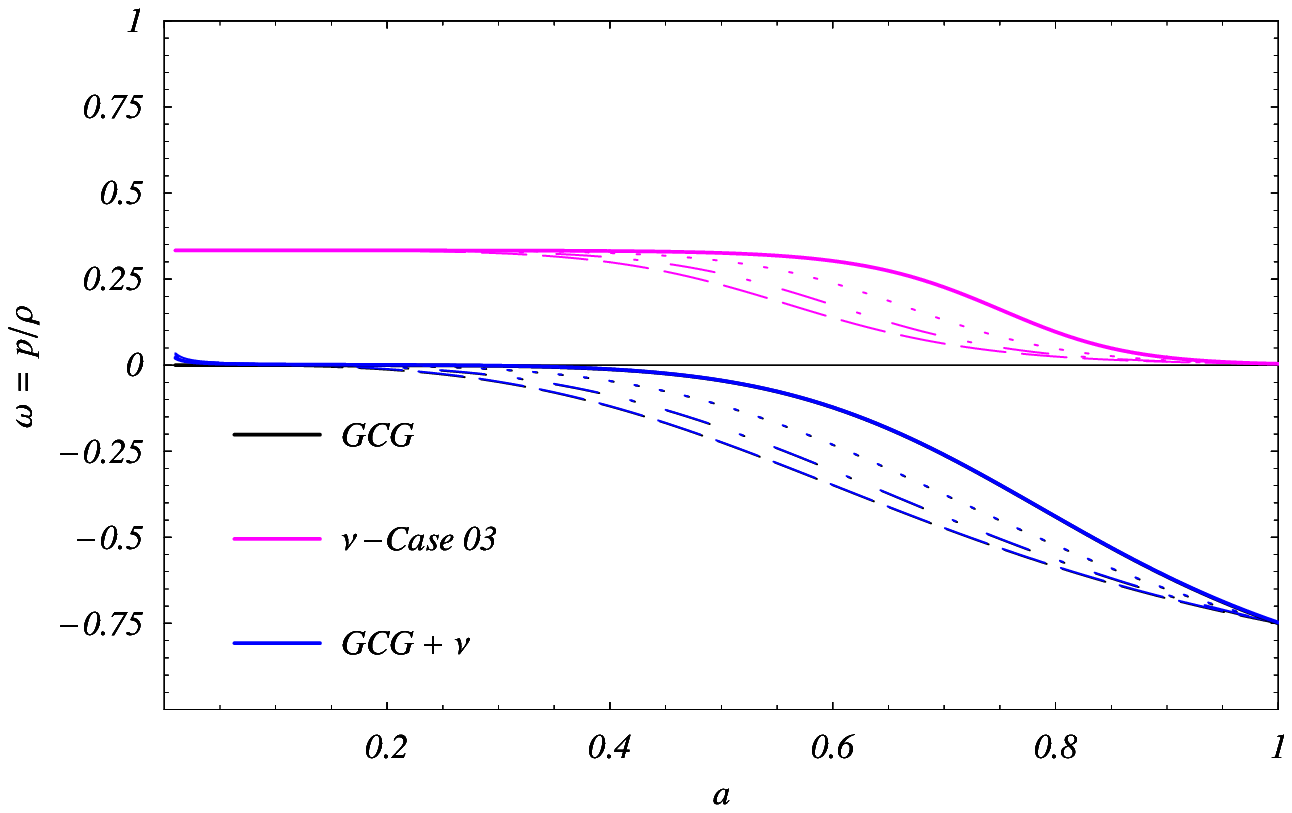,width=9.8 cm}}
\vspace{-2.5 cm}
\caption{\small Equation of state $w = p/\rho$ for 03 cosmological background scenarios in dependence on $a$.
Notice the separate behavior of the neutrino fluid (magenta) and its influence on the equation of state for the GCG scenario
Equations of state for decoupled GCG (black) and for GCG coupled to neutrinos (blue) are compared for each one of the 03 cases in correspondence with Fig.~\ref{GCG-0}.
The perturbative influence of neutrinos on the GCG background scenario is negligible.
The results are for a GCG scenario with $A_{\s}= 3/4$ and parameters $\alpha = 1$ (solid line), $1/2$ (dot line), $1/4$ (dash dot line), and $1/8$ (dash line).}
\label{GCG-C}
\end{figure}

In the Fig.~\ref{GCG-C} we verify how the energy density $\rho$ and the corresponding equations of state $w = p/\rho$ for the composed fluid deviate from the effective GCG scenario.
For mass varying CDM coupled with dark energy with $p_{\phi} = -\rho_{\phi}$, the effective GCC leads to similar predictions for $w$, independently of the scale factor $a$.
The same is not true for HDM which, in the DFG approach, while weakly coupling with dark energy, leads to the same behavior of the GCG just at late times ($a \sim 1$).

\section{Analysis and Results}

The possibility of adiabatic instabilities in cosmological scenarios was previously pointed out \cite{Afs05} in a context of a mass varying neutrino model of dark energy.
In opposition, in the usual treatment where dark matter are just coupled to dark energy, cosmic expansion together with the gravitational drag due to cold dark matter have a major impact on the stability of the cosmological background fluid.
Usually, for a general fluid for which we know the equation of state, the dominant effect on the sound speed squared $c_{s}^{\2}$ arises from the dark sector component and not from the neutrino component.

For the models where the adiabatic regime (cf. Eq.~(\ref{pap09})) implies an equation of state reproducing cosmological constant, $\Lambda$, effects, $ p_{\phi} = - \rho_{\phi}$, one obtains $c_{s}^{\2} = -1$ from the very start of the analysis.
The effective GCG is free from this inconsistency.
A detailed quantitative analysis of the stability conditions for the GCG in terms of the squared speed of sound is discussed in Ref. \cite{Ber04}.
Positive $c_{s}^{\2}$ implies that $0 \leq \alpha \leq 1$ in the GCG equation of state.
The coupling of the dark energy component with dynamical dark matter is responsible for removing such inconsistency by setting $c_{s}^{\2} \simeq \frac{d p_{\phi}}{d\rho_{\phi}} > 0$.

\begin{figure}
\vspace{-3 cm}
\centerline{\psfig{file= 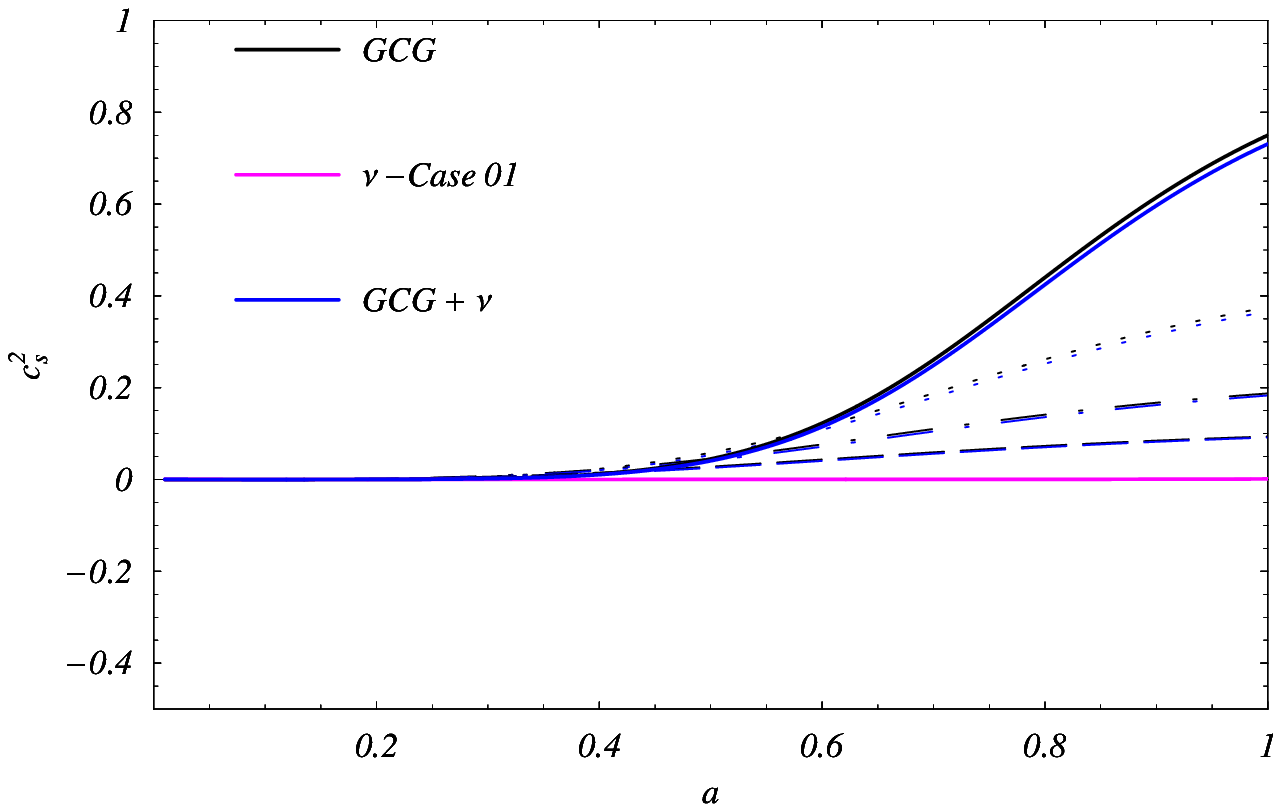,width=9.8 cm}}
\vspace{-4 cm}
\centerline{\psfig{file= 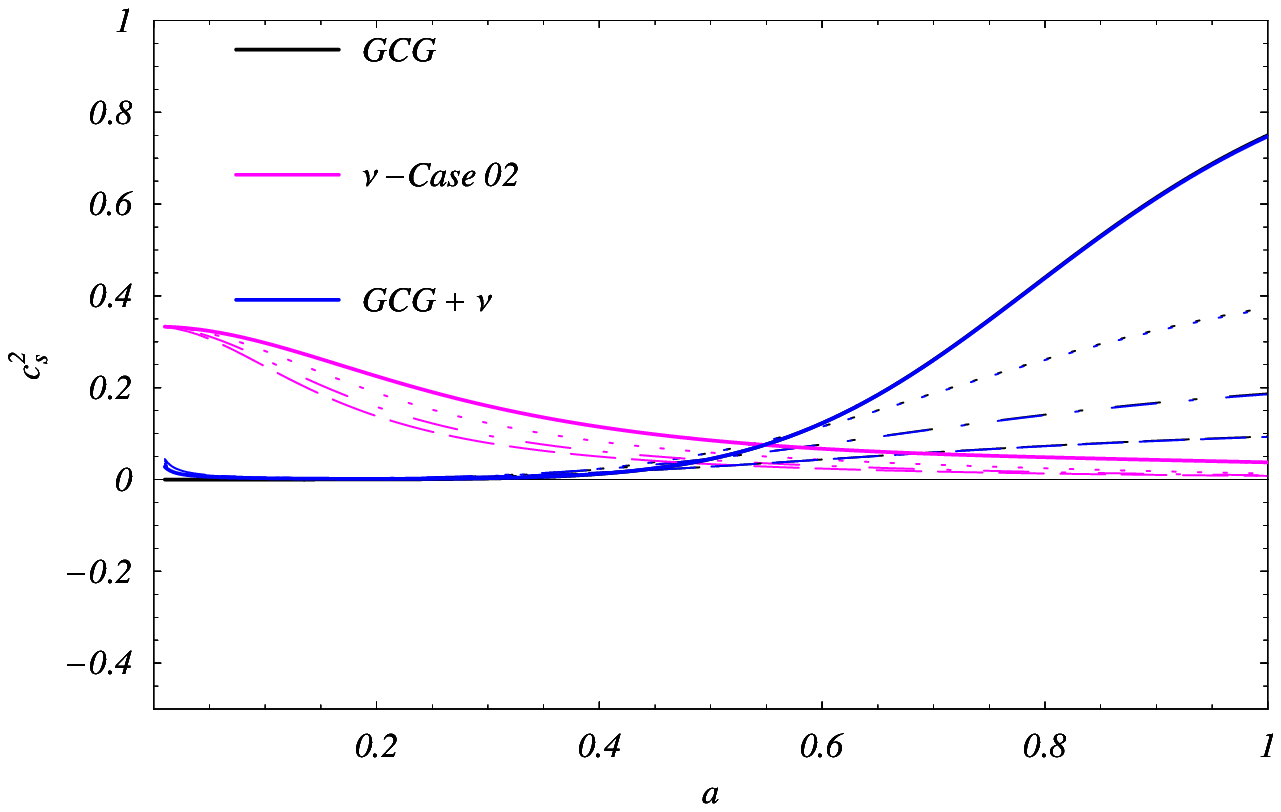,width=9.8 cm}}
\vspace{-4 cm}
\centerline{\psfig{file= 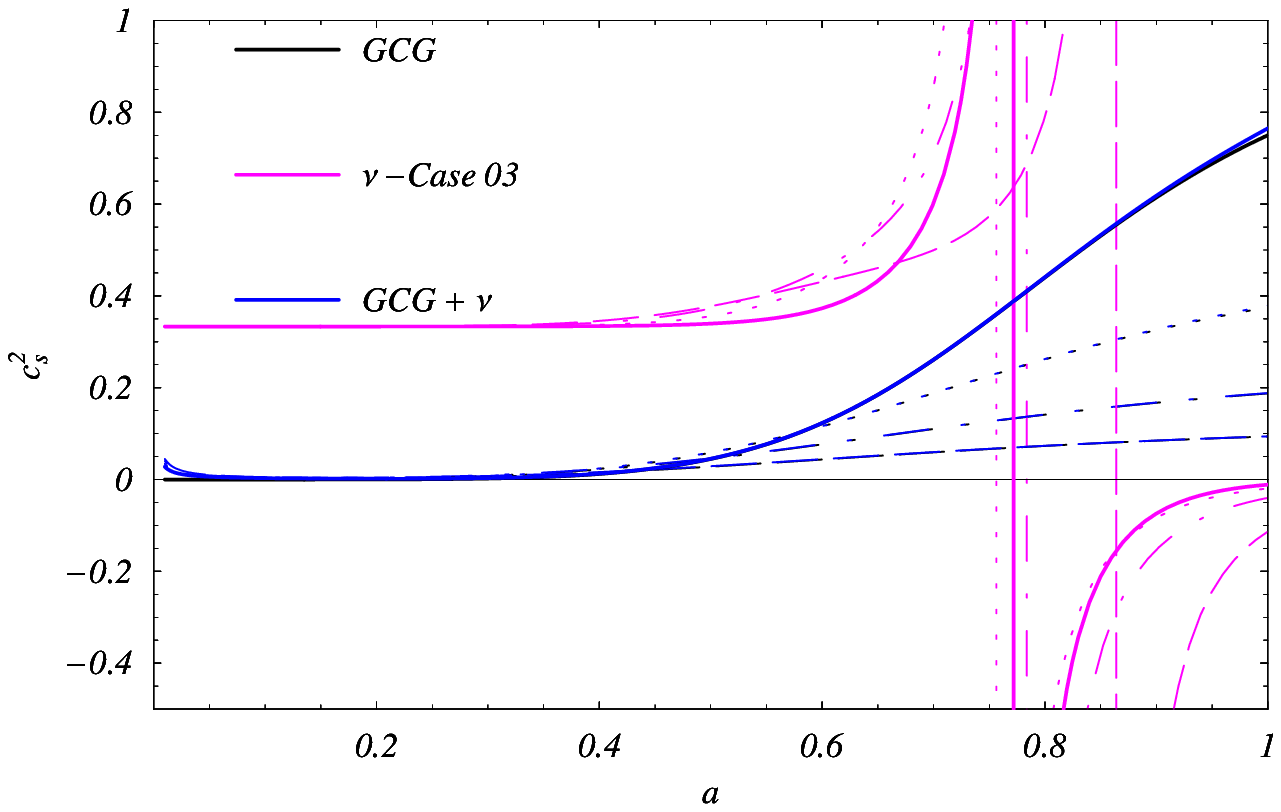,width=9.8 cm}}
\vspace{-2.5 cm}
\caption{\small Squared speed of sound $c^{\2}_{s} = \mbox{d}p/\mbox{d}\rho$ for 03 cosmological background scenarios in dependence on $a$.
Notice the results for the decoupled neutrino fluid (magenta).
The perturbative influence of neutrinos on the positiveness of $c^{\2}_{s}$ has its magnitude measured by comparing the curves for the GCG scenario with (blue) and and without (black) neutrino couplings for each one of the 03 cases in correspondence with Fig.~\ref{GCG-0}.
Even for unstable neutrino scenarios (third plot), the perturbative influence on the GCG background scenario is negligible.
The results are for a GCG scenario with $A_{\s}= 3/4$ and parameters $\alpha = 1$ (solid line), $1/2$ (dot line), $1/4$ (dash dot line), and $1/8$ (dash line).}
\label{GCG-D}
\end{figure}

In the Fig.~\ref{GCG-D} we show the results for $c^{\2}_{s}$ for a cosmological background scenario corresponding to the sum of the energy components of an effective GCG fluid: mass varying dark matter (DM), cosmon-{\em like} dark energy (DE) and perturbative DFG of neutrinos ($\nu$).
We have considered the GCG phenomenological given by $A_{\s}= 3/4$ and $\alpha = 1$ (solid line), $1/2$ (dot line), $1/4$ (dash dot line), and $1/8$ (dash line).
The perturbative influence of neutrinos on the positiveness of $c^{\2}_{s}$ has its magnitude measured by comparing the results with those for the GCG scenario with and and without neutrino perturbation for each one of the three cases in correspondence with Fig.~\ref{GCG-0}.
In this scenario, the most important result illustrated by Fig.~\ref{GCG-D} is that growing neutrino masses parameterized by $\phi^{\2}/M\bb{\phi}$ (case 03) results in instabilities at present times.

\begin{figure}
\vspace{-3 cm}
\centerline{\psfig{file= 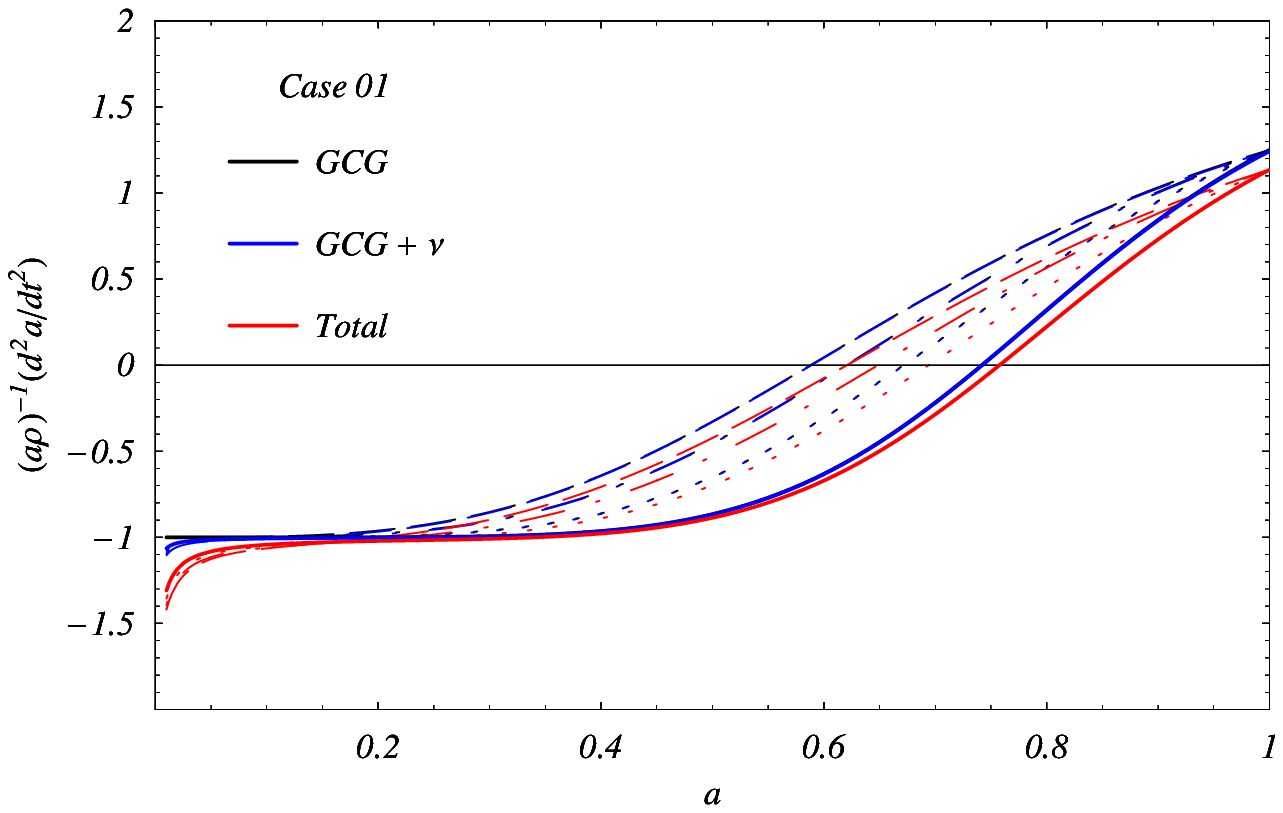,width=10 cm}}
\vspace{-3 cm}
\centerline{\psfig{file= 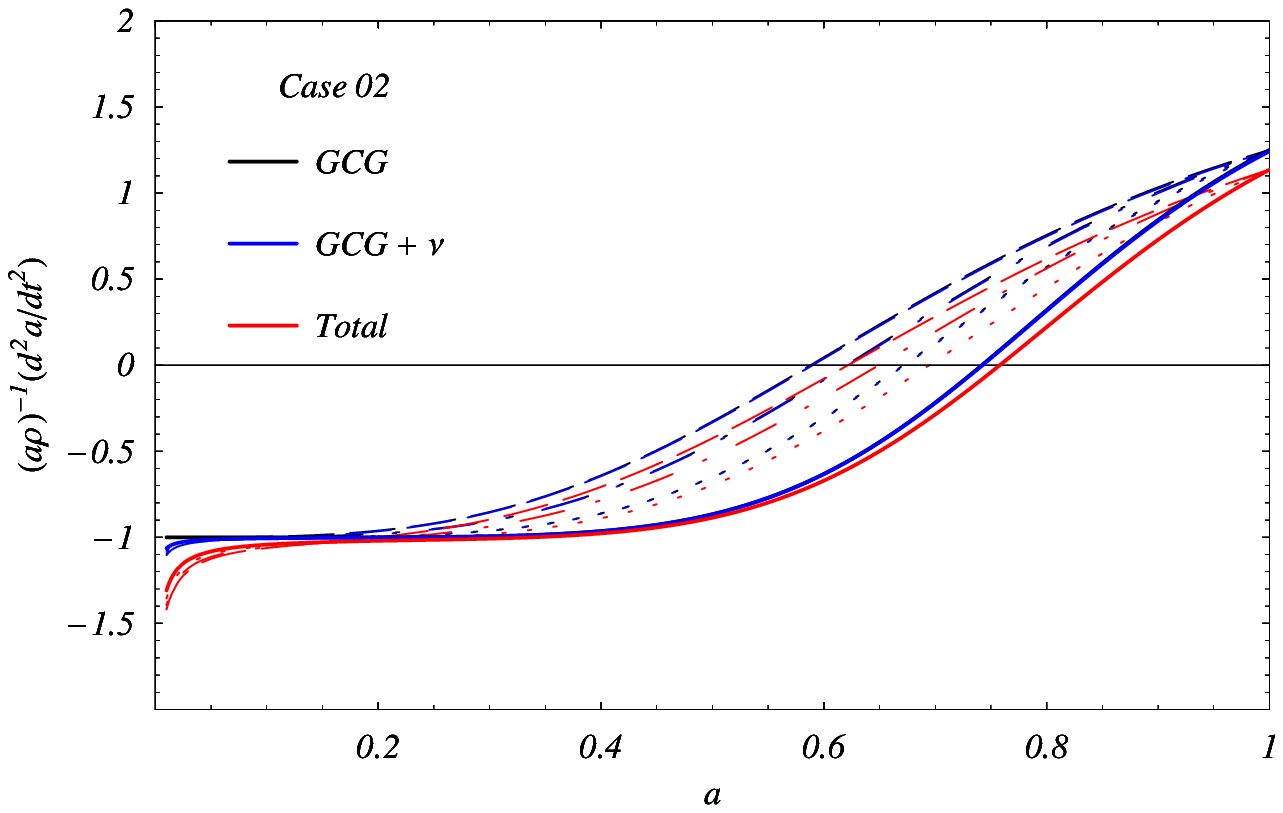,width=10 cm}}
\vspace{-3 cm}
\centerline{\psfig{file= 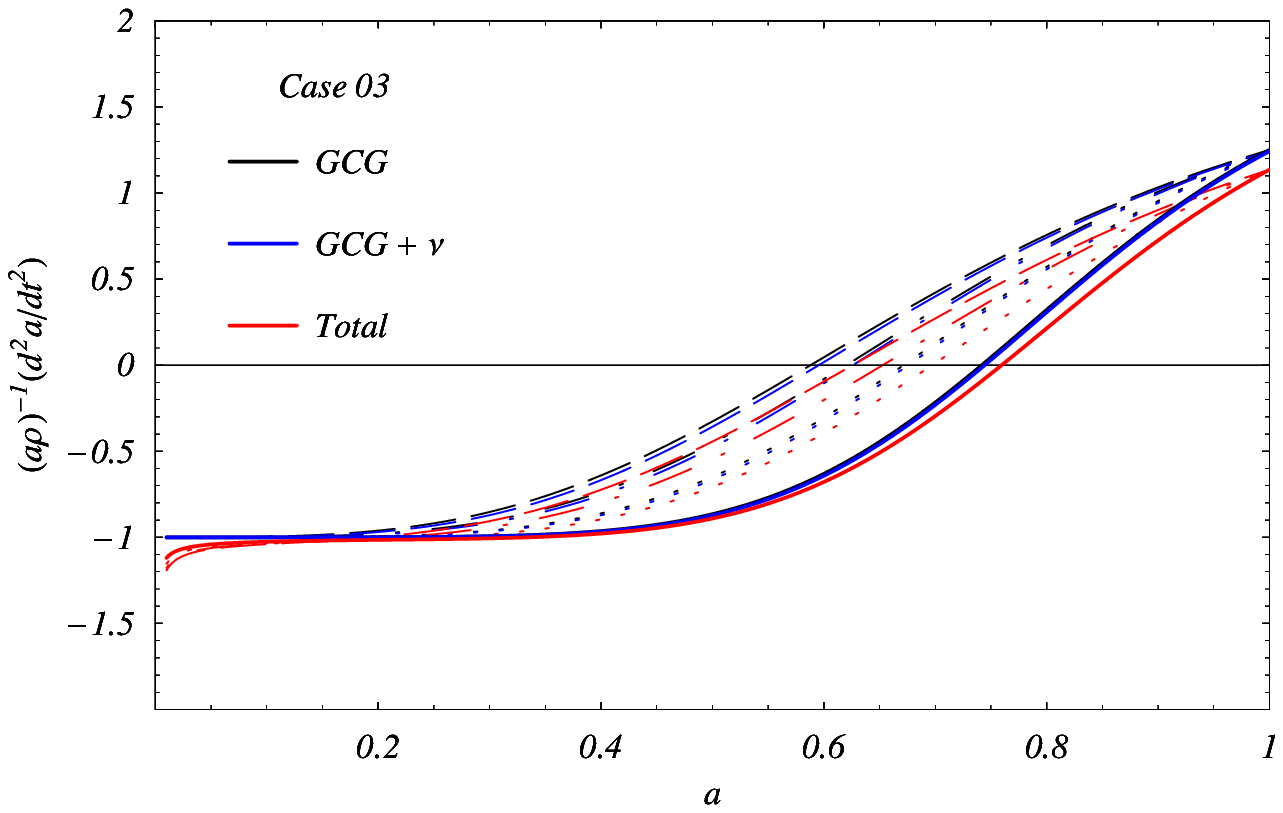,width=10 cm}}
\vspace{-2 cm}
\caption{\small Cosmic acceleration rate $\ddot{a}/(a \rho)$ for 03 cosmological background scenarios in dependence on $a$.
We have considered the GCG (black), the GCG plus neutrinos (blue), and the GCG plus neutrinos, baryons and radiation (red).
Notice that the cosmic acceleration is essentially driven by the dark sector corresponding to the GCG.
The perturbative influence of neutrinos on the cosmic acceleration is not relevant.
We use the same GCG parameters from Fig.~\ref{GCG-D}.
}
\label{GCG-E}
\end{figure}

This instability is characterized by a negative squared speed of sound for an effective coupled neutrino/dark energy fluid, and results in the exponential growth of small scale modes \cite{Bea08}.
A natural interpretation for this is that the Universe becomes inhomogeneous with the neutrinos forming denser structures or lumps \cite{Wet08,Tet08,Ber09A}.
Effectively, the scalar field could mediate an attractive force between neutrinos leading to the formation of neutrino nuggets.
Anyway, these results are consistent with the accelerated expansion of the universe ruled by the dynamical masses of CDM and neutrinos since positive values for $(1 + 3 (p_{\phi} + p_m)/(\rho_{\phi} + \rho_m))$ are observed, as we can notice in the Fig.~\ref{GCG-E}.

\section{Conclusions}

Growing neutrinos coupled to mass varying dark matter plus cosmon-{\em like} dark energy were studied assuming that the cosmological background unified fluid presents an effective behavior similar to that of the GCG.
The essential ingredient of this class of models is a neutrino mass that, through the {\em seesaw} mechanism, depends on the cosmon field and grows in the course of the cosmological evolution.
Our setup is an effective GCG decomposed into two interacting components \cite{Ber09}.
The first one behaves like mass varying CDM since it is pressure-less.
The second one corresponds to a cosmon-{\em like} dark energy component $\phi$ with equation of state given by $p_{\phi}\bb{\phi} = -\rho_{\phi}\bb{\phi}$.
Apparently the model does not look different from the interacting quintessence models where one has two different interacting fluids.
The present fraction of dark energy is set by a dynamical mechanism.
As soon as dark matter become NR, their coupling to the cosmon triggers an effective stop (or substantial slowing) of the evolution of the cosmon.
Before this event, the quintessence field follows the tracking of the minimum of an effective potential, in a kind of stationary condition which drives the adiabatic regime, and for which the coupling strength is strong compared to gravitational strength.

In the scope of finding a natural explanation for the cosmic acceleration and the corresponding adequation to stability conditions, we have proposed a systematic procedure to treat variations of fundamental parameters of neutrino cosmology ($\phi$, $m$\bb{\phi} and $M$\bb{\phi}) independent of any particular theoretical model which enforces restrictive relations among these parameters.
We have quantified the perturbative influence of such variations on the calculation of the squared speed of sound and on the cosmic acceleration of the universe, noticing that the former quantity is much more sensible to variations with respect to the exact GCG background cosmology.

An interesting feature is however that the GCG has its cosmological evolution reproduced by a CDM mass varying mechanism consistent with the mass generation mechanism for neutrinos.
Most of dark sector models predict an entangled mixture of interacting dark matter and dark energy.
We have consistently introduced the possibility of such a mixed coupling with active and sterile neutrinos, for which we have tested three hypothesis for {\em seesaw} masses.

At our approach, the mass scale $M$ and scalar field $\phi$ are tightly coupled together and evolve as a unique effective fluid.
The effective potential driving the evolution of the scalar field is decomposed into a sum of two terms, one arising from the original quintessence potential $U\bb{\phi}$, and the other from the dynamical mass of the dark matter.
The sterile neutrino mass produces the mass varying dark matter effects which indirectly result in MaVaN's, and the active neutrino has its mass perturbatively computed via {\em seesaw} relations.
Since the results for the cosmic acceleration followed by cosmological stability for GCG scenarios are well-defined, one can assert that the increase of neutrino mass acts as a cosmological clock or trigger for the crossover to the effective scenario here studied.
In fact, a scalar field associated to dark energy in connection with the SM neutrinos and the electroweak interactions may bring important insights on the physics beyond the SM \cite{Ber09C,Ber09D,Ber09E}.
The case of cosmological neutrinos, in particular, is a fascinating example where salient questions concerning SM particle phenomenology can be addressed and hopefully better understood.

\begin{acknowledgments}
A. E. B. would like to thank for the financial support from the Brazilian Agencies FAPESP (grant 08/50671-0) and CNPq (grant 300627/2007-6).
\end{acknowledgments}

\pagebreak
\newpage

\end{document}